# Electron correlation by polarization of interacting densities

Jerry L. Whitten

Department of Chemistry
North Carolina State University
Raleigh, NC 27695 USA
email: whitten@ncsu.edu

## Abstract

Coulomb interactions that occur in electronic structure calculations are correlated by allowing basis function components of the interacting densities to polarize dynamically, thereby reducing the magnitude of the interaction. Exchange integrals of molecular orbitals are not correlated. The modified Coulomb interactions are used in single-determinant or configuration interaction calculations. The objective is to account for dynamical correlation effects without explicitly introducing higher spherical harmonic functions into the molecular orbital basis. Molecular orbital densities are decomposed into a distribution of spherical components that conserve the charge and each of the interacting components is considered as a two-electron wavefunction embedded in the system acted on by an average field Hamiltonian plus $r_{12}^{-1}$. A method of avoiding redundancy is described. Applications to atoms, negative ions and molecules representing different types of bonding and spin states are discussed.

## Introduction

The description of many-electron systems by configuration interaction is a practical way to address complex systems providing the problem can be reduced to a manageable size. There is a vast literature on ways to do this ranging from perturbation methods that generate configurations and evaluate energies efficiently to methods for partitioning large systems into localized electronic subspaces or ways to balance errors in systems that are being compared.[1-11] Relatively few configurations are required to dissociate molecules correctly or to create proper spin states. However, dynamical correlation effects, particularly those associated with angular correlation, require higher spherical harmonic basis functions and this leads to a rapid increase in number of interacting configurations. In contrast, density functional methods do not rely on increasing the complexity of the wavefunction to achieve the requisite accuracy, but instead distribute the exchange-correlation effects over the density.

In the present work, we discuss a way to include dynamical correlations by introducing a polarization of components of densities involved in Coulomb interactions. The modified Coulomb interactions are used in single-determinant or configuration interaction calculations. Exchange and other integrals over molecular orbitals are not modified.

**Method**

We begin by recalling correlation effects known to be important in simple systems. In He, for example, important correlation effects can be accomplished by the polarizations,

$$\psi(1,2) = s(1)s(2) \rightarrow [s(1)+\lambda x_1 s(1)][s(2)-\lambda x_2 s(2)]+[s(1)-\lambda x_1 s(1)][s(2)+\lambda x_2 s(2)] + \ldots$$

to give an improved wavefunction,

$$\psi(1,2) = s(1)s(2) - \lambda^2 [x_1 x_2 + y_1 y_2 + z_1 z_2] s'(1)s'(2)$$

where the prime allows for a scale factor. In $H_2$, such polarizations greatly improve the wavefunction, particularly when applied to ionic components. In the $^1(\pi \rightarrow \pi^*)$ excited state of ethylene, ionic components $2p_A(1)2p_A(2) - 2p_B(1)2p_B(2)$, occur on nuclei A and B, and for such distributions angular correlations are important. These are well-known effects in configuration interaction applications.

Many very accurate configuration interaction calculations have been carried out on atoms, molecules and clusters using large basis sets that include higher spherical harmonic functions.[1-11] As systems increase in size, the additional functions required for angular correlation cause a rapid increase in the number of configurations, limiting the applicability of the method.

In the present work, we consider an alternative approach that avoids introducing higher spherical harmonic basis functions as explicit configurations. The argument proceeds as follows. Starting with a single determinant wavefunction and allowing excitations of spatial orbitals modifies the wavefunction and energy expression

$$\begin{array}{cccc} \varphi'_i & \varphi'_j & \varphi'_i & \varphi'_j \\ \uparrow & \uparrow & \uparrow & \uparrow \end{array}$$

$$<\psi|H|\psi> = (norm) < \det(\ldots \varphi_i(1) \ldots \varphi_j(2) \ldots ) | H | \det(\ldots \varphi_i(1) \ldots \varphi_j(2) \ldots ) >$$

$$\psi(1,2,3\ldots N) = (N!)^{-1/2} \det(\ldots \varphi_i(1)\ldots\varphi_j(2)\ldots) + \lambda(N!)^{-1/2} \det(\ldots\varphi'_i(1)\ldots\varphi'_j(2)\ldots) + \ldots$$

As noted above, excitations corresponding to a linear response of the orbitals, $\varphi'_i(1) = x_1 \varphi_i(1)$, $\varphi'_j(2) = x_2 \varphi_j(2)$, similarly for y and z, introduce angular correlation into the wavefunction and are particularly important. The energy lowering can be considered as modifying the Coulomb repulsion, $(\varphi_i(1)\varphi_i(1)|r_{12}^{-1}|\varphi_j(2)\varphi_j(2))$, and this provides a convenient way to introduce the correlation energy lowering into the formalism.

The plan is as follows:

1) Introduce $x_1x_2 + y_1y_2 + z_1z_2$, with appropriate choice of origins, into the formalism at the level of basis function interactions.

2) Use the result to modify the Coulomb interaction of molecular orbitals, keeping exchange and other interactions unchanged.

$$(\varphi_i(1)\varphi_i(1) | r_{12}^{-1} | \varphi_j(2)\varphi_j(2))_{exact} \rightarrow (\varphi_i(1)\varphi_i(1) | r_{12}^{-1} | \varphi_j(2)\varphi_j(2))_{corr}$$

3) Construct wavefunctions by configuration interaction, using the modified Coulomb interactions to evaluate energies.

The objective is to retain the flexibility of a multi-configuration description in order to describe bonding, molecular dissociation and spin states. Continuing to use configuration interaction requires that configurations generated explicitly are not the same as the excitations built into the formalism. For example, if d-functions are included in the basis for polarization purposes, they must be projected out of the virtual molecular orbital basis to avoid redundancy in p-shell correlation.

We consider the electron pair, $\varphi_i(1)\varphi_j(2)$, embedded in the remainder of the system, omitting exchange which remains exact in the method. The Coulomb interaction can be used to track excitation contributions,

$$<\varphi_i(1)\varphi_j(2)) | r_{12}^{-1} | \varphi_i(1)\varphi_j(2)) >$$

$$= \sum_{kk'mm'} (coef)_{kk'mm'}^{ij} < f_k(1) f_{k'}(2) | r_{12}^{-1} | f_m(1) f_{m'}(2) >$$

$$\downarrow \quad \downarrow \qquad \quad \downarrow \quad \downarrow$$

$$f'_k \quad f'_{k'} \qquad f'_m \quad f'_{m'}$$

However, it is complicated to carry out the four component excitations as expressed above. Instead, we expand the densities retaining only the moments that contain charge. Functions are assumed real and expressed as linear combinations of basis functions, $\varphi_i = \sum c_{ki} f_k$. The distribution of component densities over many points in space will be an important part of the argument, and we assume expansions in terms of Gaussian functions,

$f = (norm) x^p y^q z^s \exp(-ar^2)$, located at specific origins. Since the product of two spherical Gaussian functions is a new Gaussian, $F_p$, at an origin, $P$, it is always possible to resolve the product of functions containing $x^p y^q z^s$ as a moment expansion,

$$f_k f_m = <f_k | f_m> F_P^{km} F_P^{km} + F_P^{km} F_P^{km}[\lambda_x(x-P_x) + \lambda_{xy}(x-P_x)(y-P_y) + \lambda_{xx}((x-P_x)^2 - \lambda_0)...]$$

and similarly for the other density to give

$$(\varphi_i(1)\varphi_i(1) | r_{12}^{-1} | \varphi_j(2)\varphi_j(2)) =$$
$$\sum_{kk'mm'} c_{ki} c_{mi} c_{k'j} c_{m'j} <f_k|f_m><f_{k'}|f_{m'}> (F_P^{km}(1) F_Q^{k'm'}(2) | r_{12}^{-1} | F_P^{km}(1) F_Q^{k'm'}(2))$$

We refer to this expansion as the spherical component approximation. We shall use the spherical component expansion to include correlation and not as a substitute for the correct evaluation of electron repulsion interactions which would require all moments. The accuracy of the spherical component representation and the possibility of improvement by renormalization is discussed later.

To simplify the notation, let $P(1)Q(2) = F_P^{km}(1) F_Q^{k'm'}(2)$, where $P$ and $Q$ denote single spherical Gaussian functions centered at points $(P_x, P_y, P_z)$ and $(Q_x, Q_y, Q_z)$ in space with exponents $\frac{\alpha}{2}$ and $\frac{\beta}{2}$ respectively. We now treat $P(1)Q(2)$ as a two-electron wavefunction embedded in the system. We postulate an effective Hamiltonian does not change the wavefunction, $P(1)Q(2)$, and thus take it to be that for three-dimensional harmonic oscillators with potentials determined by α and β

$$(h_1 + h_2)P(1)Q(2) = E_0 P(1)Q(2) = (\tfrac{3}{2}\alpha + \tfrac{3}{2}\beta)P(1)Q(2)$$

We now add electron repulsion explicitly and allow the wavefunction to respond, solving approximately

$$(h_1 + h_2 + \tfrac{1}{r_{12}})(P(1)Q(2) + \chi(1,2)) = (E_0 + E_\delta)(P(1)Q(2) + \chi(1,2))$$

by variational energy minimization, where

$$\chi(1,2) = c\,[(x_1 - P_x)(x_2 - Q_x) + (y_1 - P_y)(y_2 - Q_y) + (z_1 - P_z)(z_2 - Q_z)]P'(1)Q'(2)$$

and the prime denotes an exponent scale factor, $\eta$, that maximizes the energy lowering. The energy lowering, $E_\delta$, can be found by solving the 2x2 variational problem or by perturbation theory. The latter gives

$$E_\delta = -\frac{|<P(1)Q(2)|r_{12}^{-1}|\chi(1,2)>|^2}{<\chi(1,2)|h_1+h_2|\chi(1,2)> - <P(1)Q(2)|h_1+h_2|P(1)Q(2)>}$$

The question of including $r_{12}^{-1}$ in the definition of the zeroth order Hamiltonian or including it explicitly depends on the value of

$$< \chi(1,2) | r_{12}^{-1} | \chi(1,2) > - < P(1)Q(2) | r_{12}^{-1} | P(1)Q(2) >$$

which turns out to be small for the optimum scale factor, $\eta$, and this contribution is not included. The expression for $E_\delta$ for spherical Gaussians $P$ and $Q$ is

$$E_\delta = - C \left(\alpha^{-1} + \beta^{-1}\right)^{-2} (\alpha + \beta)^{-2} \left[ exp\left(-(1+\eta)\left(\alpha^{-1} + \beta^{-1}\right)^{-1}\right) PQ^2 \right]$$

$$\text{where } C = 2^9 (3\pi)^{-1} \eta^5 (1+\eta)^{-7} \left[\tfrac{5}{4}(\eta + \eta^{-1}) - \tfrac{3}{2}\right]^{-1}$$

Maximizing $-E_\delta$ with respect to $\eta$ for $PQ^2$=0, gives $\eta$ =1.47, and for this choice of η

$$E_\delta = - 0.55969166 \left(\alpha^{-1} + \beta^{-1}\right)^{-2} (\alpha + \beta)^{-2} \left[ exp\left(- 2.47\left(\alpha^{-1} + \beta^{-1}\right)^{-1} PQ^2\right) \right]$$

We define, $E_\delta$, as a correlation contribution associated with the PQ term in the Coulomb interaction and create the modified integral, $(f_k(1)f_m(1) | r_{12}^{-1} | f_{k'}(2)f_{m'}(2)) + E_\delta$. This gives an additional set of basis function integrals to be used to modify Coulomb interactions of atomic and molecular orbitals. There are no adjustable parameters in the method.

A potential problem associated with excitations at the basis function level is the possibility that the excitation may overlap terms already in the expression. If $f_i, f_j$ both belong to the highest set of spherical harmonics included in the basis, there is no redundancy to a first approximation since functions generated are not present in the basis. However, in first-row atoms, $ss \rightarrow pp$, $sp \rightarrow pd$ excitations are partially blocked by the occupancy of the $p$-shell. We approximate the blocking by examining the number of channels that are open in the atom to which the basis function belongs. If the p-shell is occupied by n electrons, the correlation contributions for an s-excitation is reduced to $\frac{(6-n)}{6} E_\delta$. Since there are no d-electrons in the virtual space in the atoms and molecules considered, there is no reduction in $E_\delta$ for $pp' \rightarrow dd'$ excitations. There are ways to treat the blocking effect self-consistently, but, because the inter-shell correlation contributions are smaller, the simplest method appears to be adequate. For transition metal systems, similar blocking arguments can be made.

We consider first applications to several simple systems containing only s-type basis functions. In these cases, the expansion of the density discussed above is exact and the only question is whether the correlation construction is useful.

## Systems with only s-type basis functions

Results for He, Li, Be, Li$^{-1}$, H$^{-1}$ and H$_2$ are summarized in Table 1. Energies are given for single-determinant and CI "exact" calculations in which no additional correlation is included ($E_\delta = 0$) and for "correlated" calculations in which the $E_\delta$ contributions to atomic or molecular orbital Coulomb integrals is included. In all cases, the inclusion of the $E_\delta$ contribution is found to move the single determinant and CI total energies much closer to the experimental values.

**Table 1. Systems containing only s-type orbitals.**
Calculations are for a near Hartree-Fock basis, double zeta contraction. Energies are from SCF and CI calculations, and from calculations including E$\delta$, representing correlation contributions from p-type configurations not included in the calculations.

|  | 1-det | CI | expt[a] |  |  |
|---|---|---|---|---|---|
| **He** |  |  |  |  |  |
| (2 s-type orbitals) |  |  |  |  |  |
| SCF, CI (no E$_\delta$) | -2.8612 | -2.8771 |  |  |  |
| E$_\delta$ included | -2.8834 | -2.8994 | -2.9028 |  |  |
| **Li** |  |  |  |  |  |
| (4 s-type orbitals) |  |  |  |  |  |
| SCF, CI (no E$_\delta$) | -7.4325 | -7.4471 |  |  |  |
| E$_\delta$ included | -7.4554 | -7.4709 | -7.4762 |  |  |
| **Be** |  |  |  |  |  |
| 4 s-type orbitals) |  |  |  |  |  |
| SCF, CI (no E$_\delta$) | -14.5685 | -14.5895 |  |  |  |
| E$_\delta$ included | -14.6237 | -14.6445 | -14.6649 |  |  |

|  | 1-det | CI | expt[a] | electron affinity calc | expt[e] |
|---|---|---|---|---|---|
| **H-minus**[b] |  |  |  |  |  |
| SCF, CI (no E$_\delta$) | -0.4836 | -0.5135 |  |  |  |
| E$_\delta$ included | -0.4997 | -0.5297 | -0.5277 | 0.81 | 0.75 |
| **Li-minus**[b] |  |  |  |  |  |
| SCF, CI (no E$_\delta$) | -7.4265 | -7.4504 |  |  |  |
| E$_\delta$ included | -7.4650 | -7.4885 | -7.4989 | 0.50 | 0.62 |

|  | 1-det | CI | expt[a] | dissoc energy (eV) calc[c] | expt[c] |
|---|---|---|---|---|---|
| **H$_2$** |  |  |  |  |  |
| (4 s-type orbitals) |  |  |  |  |  |
| SCF, CI (no E$_\delta$) | -1.1284 | -1.1538 |  |  |  |
| E$_\delta$ included | -1.1439 | -1.1694 | -1.1742 | 4.61 |  |
| (5s-type orbitals, including midpoint function) |  |  |  |  |  |
| SCF, CI (no E$_\delta$) | -1.1311 | -1.1569 |  |  |  |
| E$_\delta$ included | -1.1475 | -1.1735 | -1.1742 | 4.72 | 4.75 |

[a]Total energies are in Hartree a.u., 1 a.u.=27.21 eV.; experimental values are from Ref. 12, 13.
[b]Basis is reoptimized for the negative ions.
[c]The calculated equilibrium distance is the same as experiment, 0.74 Å.

Next, we consider application to atoms B through Ne. Since p-type basis functions are present, the spherical density component (leading term) approximation is a simplification of uncertain validity. We consider a renormalization to improve the accuracy. Specifically, two ways to incorporate $E_\delta$ have been considered for each atom or molecule investigated:

1) Direct modification of integrals (the spherical density component approximation)

$(\varphi_i(1)\varphi_i(1) | r_{12}^{-1} | \varphi_j(2)\varphi_j(2))$ is replaced by $(\varphi_i(1)\varphi_i(1) | r_{12}^{-1} | \varphi_j(2)\varphi_j(2)) + \tilde{E}_\delta$

where $\tilde{E}_\delta$ is the change in the atomic or molecular orbital integral due to the $E_\delta$ contribution at the basis function level.

2) Renormalization at the atomic or molecular orbital level.

$(\varphi_i(1)\varphi_i(1) | r_{12}^{-1} | \varphi_j(2)\varphi_j(2))$ is replaced by

$$(\varphi_i(1)\varphi_i(1) | r_{12}^{-1} | \varphi_j(2)\varphi_j(2)) + \tilde{E}_\delta \frac{(\varphi_i(1)\varphi_i(1)| r_{12}^{-1} |\varphi_j(2)\varphi_j(2))}{(\varphi_i(1)\varphi_i(1)| r_{12}^{-1} |\varphi_j(2)\varphi_j(2))_{spher}}$$

where *spher* denotes the value calculated using only spherical components of the interacting densities. There is no change from 1) if the basis functions are all s-type.

**Atoms B through Ne**

The basis for each atom is a near Hartree Fock set of atomic orbitals plus extra two-component s- and p-type functions consisting of the two smaller exponent components of the Hartree-Fock atomic orbital. An additional shorter range 2-component s-type orbital is added to provide improved radial correlation of the 1s orbital. The basis can be described as 1s(10), 2s(5), 2p(5), 2s′(2), 2p′(2), 1s′(2), a total of ten contracted functions with the total number of Gaussian functions for each orbital given in parentheses. Energies with and without $E_\delta$ contributions are reported in Table 2. Also included in the table are literature values from work by Dunning[3] and Sasaki and Yoshimine[4]. The former results are for a "correlation consistent" basis set that is designed for atoms and molecules, while the latter results are for a very large basis CI calculation. We note that including $E_\delta$ lowers the energy substantially at both the single determinant and CI levels. The calculated total energies lie between the correlation-consistent and very large CI values.  Although the simplest treatment with no renormalization gives lower energies, this is likely due to overestimates of the correlation by the spherical component approximation.  This question will be discussed in a later section.  It should be noted that these are relatively small basis set calculations and there are no adjustable parameters in the different applications.  Values for electron affinities are exceptionally good for C and O and fairly good for F.  In Table 3, results are reported for a basis slightly reduced in size to 9 functions by removal of the 1s′(2). These results provide reference energies for diatomic molecule calculations where excitations from the 1s shell are not allowed.

**Table 2.  Atoms B - Ne and negative ions.**
Calculations are for a near Hartree-Fock basis: double zeta contraction, 10 basis orbitals.
The 1s orbital is included in the CI.   Energies are given for normal SCF and CI calculations
Variational  calculations using high quality basis sets are included for comparison.

|  | 1-det[a] | CI | CI cc-basis[b] | CI vl-basis[c] |
|---|---|---|---|---|
| **B** | | | | |
| SCF, CI  (no $E_\delta$) | -24.5284 | -24.5667 | | |
| $E_\delta$ included (no renorm) | -24.5735 | -24.6118 | | |
| $E_\delta$ included (renorm) | -24.5708 | -24.6086 | -24.5982 | -24.6500 |
| **C** | | | | |
| SCF, CI  (no $E_\delta$) | -37.6882 | -37.7422 | | |
| $E_\delta$ included (no renorm) | -37.7640 | -37.8185 | | |
| $E_\delta$ included (renorm) | -37.7552 | -37.8093 | -37.7796 | -37.8393 |
| **N** | | | | |
| SCF, CI  (no $E_\delta$) | -54.4001 | -54.4536 | | |
| $E_\delta$ included (no renorm) | -54.4924 | -54.5458 | | |
| $E_\delta$ included (renorm) | -54.4762 | -54.5296 | -54.5118 | -54.5812 |
| **O** | | | | |
| SCF, CI  (no $E_\delta$) | -74.8068 | -74.8858 | | |
| $E_\delta$ included (no renorm) | -74.9302 | -75.0089 | | |
| $E_\delta$ included (renorm) | -74.9034 | -74.9821 | -74.9685 | -75.0542 |
| **F** | | | | |
| SCF, CI  (no $E_\delta$) | -99.4085 | -99.5185 | | |
| $E_\delta$ included (no renorm) | -99.5671 | -99.6765 | | |
| $E_\delta$ included (renorm) | -99.5280 | -99.6375 | -99.6122 | -99.7166 |
| **Ne** | | | | |
| SCF, CI  (no $E_\delta$) | -128.5459 | -128.6876 | | |
| $E_\delta$ included (no renorm) | -128.7441 | -128.8851 | | |
| $E_\delta$ included (renorm) | -128.6908 | -128.8320 | -128.7919 | -128.9168 |

|  |  |  | electron affinity (eV) | |
| --- | --- | --- | --- | --- |
|  |  |  | calc | expt[e] |
| **C-minus[d]** | | | | |
| SCF, CI (no $E_\delta$) | -37.7059 | -37.7663 | | |
| $E_\delta$ included (renorm) | -37.7965 | -37.8560 | 1.27 | 1.26 |
| **O-minus[d]** | | | | |
| SCF, CI (no $E_\delta$) | -74.7790 | -74.9053 | | |
| $E_\delta$ included (renorm) | -74.9127 | -75.0382 | 1.53 | 1.46 |
| **F-minus[d]** | | | | |
| SCF, CI (no $E_\delta$) | -99.4460 | -99.5904 | | |
| $E_\delta$ included (renorm) | -99.6090 | -99.7519 | 3.10 | 3.40 |

[a] Total energies are in Hartree a.u., 1.a.u. = 27.21 eV.
[b] Dunning, Ref 3., correlation consistent basis (10s5p2d1f)/[4s3p2d1f].
[c] Sasaki, Yoshimine, Ref. 4., very large basis, nearly full-CI
[d] Basis is reoptimized for the negative ions.
[e] Ref. 13.

**Table 3. Atoms B - Ne. Reference energies for molecular dissociation energy calculations.**
Calculations are for a near Hartree-Fock basis: double zeta contraction, 10 basis orbitals;
The 1s orbital is not included in the CI. Energies are given for normal SCF and CI calculations
and for calculations that include Eδ.

|  | 1-det[a] | CI |
|---|---|---|
| **B** | | |
| SCF, CI (no Eδ) | -24.5284 | -24.5667 |
| $E_\delta$ included (no renorm) | -24.5735 | -24.6118 |
| Eδ included (renorm) | -24.5708 | -24.6086 |
| **C** | | |
| SCF, CI (no Eδ) | -37.6882 | -37.7256 |
| $E_\delta$ included (no renorm) | -37.7488 | -37.7866 |
| Eδ included (renorm) | -37.7414 | -37.7788 |
| **N** | | |
| SCF, CI (no Eδ) | -54.4001 | -54.4356 |
| $E_\delta$ included (no renorm) | -54.4807 | -54.5161 |
| Eδ included (renorm) | -54.4662 | -54.5016 |
| **O** | | |
| SCF, CI (no Eδ) | -74.8068 | -74.8712 |
| $E_\delta$ included (no renorm) | -74.9224 | -74.9865 |
| Eδ included (renorm) | -74.8971 | -74.9613 |
| **F** | | |
| SCF, CI (no Eδ) | -99.4085 | -99.5012 |
| $E_\delta$ included (no renorm) | -99.5631 | -99.6552 |
| Eδ included (renorm) | -99.5249 | -99.6171 |
| **Ne** | | |
| SCF, CI (no Eδ) | -128.5459 | -128.6692 |
| $E_\delta$ included (no renorm) | -128.7441 | -128.8666 |
| Eδ included (renorm) | -128.6908 | -128.8135 |

[a] Total energies are in Hartree a.u., 1.a.u. = 27.21 eV.

**Diatomic molecules**

Applications to diatomic molecules representing different types of bonding and spin states are reported in Table 4. These calculations, because of changes in spin states in the atoms and molecules and the different types of bonding, provide a good test of the method. The basis is exactly the same as described above except for the addition of a set of five two-component d-orbitals on each nucleus. These d-type orbitals are optimized at the Hartree-Fock level to allow polarization effects associated with bond formation. Since the purpose of the present work is to avoid including d-type orbitals in the CI calculations, we want to eliminate d-orbitals from the virtual molecular orbital basis. This is accomplished by a unitary transformation of the virtual orbitals to maximize their overlap with the basis set that excludes d-type functions. The result is a virtual basis largely free of d-orbital contributions. If this step is ignored, there is a partial double counting of correlation contributions by inclusion of p to d excitations in the CI and this leads to larger dissociation energies. Including the correlation contribution, $E_\delta$, is found to improve dissociation energies substantially compared to CI calculations using the same basis. Calculated equilibrium internuclear distances are close to the experimental values except for $F_2$ which is longer by 0.05 Å and the energy varies slowly near the minimum. Although similar results are obtained for the two methods of incorporating $E_\delta$, the best dissociation energies are obtained for renormalization of the interacting molecular orbital densities.

**Table 4. Molecular calculations.**
Total energies, variation with internuclear distance and dissociation energies. Calculations are for a near atomic Hartree-Fock basis: double zeta contraction. The same basis was used for the atomic energies in Table 3. Energies are given for normal SCF and CI calculations and for calculations that include E$\delta$.

|  | 1-det[a] | CI | Dissociation energy (eV) | |
|---|---|---|---|---|
|  |  |  | CI | Expt[b] |
| **O2** |  |  |  |  |
| SCF, CI (no E$_\delta$) | -149.6499 | -149.9025 | 4.36 |  |
| E$\delta$ included (no renorm) | -149.8915 | -150.1478 | 4.76 |  |
| E$\delta$ included (renorm) | -149.8496 | -150.1067 | **5.01** | **5.21** |
|  | R (bohr) | E |  |  |
|  | 2.180 | -150.0988 |  |  |
|  | 2.28* | -150.1046 |  |  |
|  | 2.285 | -150.1047 |  |  |
|  | 2.380 | -150.1023 |  |  |
| **CO** |  |  |  |  |
| SCF, CI (no E$_\delta$) | -112.7784 | -112.9893 | 10.68 |  |
| E$\delta$ included (no renorm) | -112.9764 | -113.1875 | 11.28 |  |
| E$\delta$ included (renorm) | -112.9425 | -113.1546 | **11.28** | **11.23** |
|  | R (bohr) | E |  |  |
|  | 2.030 | -113.1492 |  |  |
|  | 2.13* | -113.1546 |  |  |
|  | 2.230 | -113.1469 |  |  |
| **F2** |  |  |  |  |
| SCF, CI (no E$_\delta$) | -198.7591 | -199.0369 | 0.94 |  |
| E$\delta$ included (no renorm) | -199.0746 | -199.3551 | 1.22 |  |
| E$\delta$ included (renorm) | -199.0134 | -199.2934 | **1.61** | **1.66** |
|  | R (bohr) | E |  |  |
|  | 2.568 | -199.2890 |  |  |
|  | 2.668* | -199.2934 |  |  |
|  | 2.718 | -199.2943 |  |  |
|  | 2.738 | -199.2944 |  |  |
|  | 2.768 | -199.2945 | 1.64eV |  |
|  | 2.818 | -199.2942 |  |  |

|   |   | 2.868 | -199.2934 |   |   |
|---|---|---|---|---|---|
| **NO** |   |   |   |   |   |
|   |   | -129.2818 | -129.5162 | 5.70 |   |
| SCF, CI (no E$_\delta$) |   | -129.4985 | -129.7348 | 6.32 |   |
| E$\delta$ included (no renorm) |   | -129.4620 | -129.6982 | **6.40** | **6.61** |
| E$\delta$ included (renorm) |   |   |   |   |   |
|   |   | R (bohr) | E |   |   |
|   |   | 2.075 | -129.6905 |   |   |
|   |   | 2.175* | -129.6982 |   |   |
|   |   | 2.275 | -129.6939 |   |   |
| **N2** |   |   |   |   |   |
| SCF, CI (no E$_\delta$) |   | -108.9786 | -109.2010 | 8.98 |   |
| E$\delta$ included (no renorm) |   | -109.1706 | -109.3945 | 9.86 |   |
| E$\delta$ included (renorm) |   | -109.1398 | -109.3645 | **9.83** | **9.90** |
|   |   | R (bohr) | E |   |   |
|   |   | 1.974 | -109.3554 |   |   |
|   |   | 2.074* | -109.3645 |   |   |
|   |   | 2.174 | -109.3583 |   |   |

[a]Total energies are in Hartree a.u., 1 a.u. = 27.21 eV.
[b]Refs. 14-15. Experimental internuclear distances are denoted by *.

**Other considerations**

An important requirement in the calculation of dissociation energies is a consistent treatment of electron spin in atoms and molecules where the atom often has a higher spin state. The intent in the present work is to apply the added correlation, $E_\delta$, to electrons with opposite spin and to rely on the exchange hole for same spin electrons. The inclusion of $E_\delta$, is referenced to electron pairs $\varphi_i(1)\varphi_j(2)$, however, and this results in the correct correlation contribution for electrons with opposite spin in closed shell systems, but leaves a residual $E_\delta$ correlation, reduced by a factor of two, for interactions involving singly occupied orbitals. This slightly overestimates the correlation contribution to electrons with the same spin. Proceeding in this way has the advantage of preserving the degeneracy of different $m_s$ components of a spin state.

An important question is the accuracy of the representation of a molecular orbital density by the spherical component approximation since this is related to the value of the calculated correlation contribution. We shall use $O_2$ and chlorophyll-a to illustrate factors that may affect the accuracy. For $O_2$, we consider the 16 molecular orbitals that result after transforming out d-type orbitals from the virtual space and excluding the two oxygen 1s atomic orbitals. This gives a total of 136 Coulomb integrals for the small $O_2$ basis used in Table 4. In Fig. 1, upper graph, the difference between correlated (no renormalization) and exact values of $(\varphi_i(1)\varphi_i(1)|r_{12}^{-1}|\varphi_j(2)\varphi_j(2))$ $i \leq j$ is plotted vs. integral number, 1-136. The first observation is that the effect of including $E_\delta$ is small except for several spikes in the graph. These spikes can be traced to interactions with a single virtual orbital. In Fig.1, lower panel, the values of $(\varphi_i(1)\varphi_i(1)|r_{12}^{-1}|\varphi_j(2)\varphi_j(2))_{exact}$ and the value of the integral calculated using the spherical component approximation, $(\varphi_i(1)\varphi_i(1)|r_{12}^{-1}|\varphi_j(2)\varphi_j(2))_{spher}$ are plotted. The latter graph shows that the spherical approximation breaks down at the same points noted in the difference graph. If we renormalize the densities in the Coulomb integral as described earlier, the spikes are greatly reduced in the new difference plot as shown in the figure. For this particular system, the problematic integrals involve interaction with a specific virtual orbital and would have the effect of reducing the energy of an excited configuration. In $O_2$, the error even without renormalization turns out to be small as seen in Table 4. In other systems, however, there may be a near linear dependency of the basis and the errors could be much larger, particularly affecting high energy virtual orbitals, and it would be important to detect this over-correlation error in advance. The above discussion identifies two ways to do this using information already available.

Fig. 2, gives the same information for a calculation on chlorophyll-a where 10 molecular orbitals below the highest occupied were selected along with the lower energy 16 orbitals from the virtual space. The graphs refer to the resulting 351 molecular orbital integrals
$(\varphi_i(1)\varphi_i(1)|r_{12}^{-1}|\varphi_j(2)\varphi_j(2))$ $i \leq j$. In this system, where the molecular orbitals are extensively delocalized, the Coulomb integrals are smaller and the spherical component approximation is found to be closer to the exact integrals (lower graphs) than in the more spatially localized $O_2$ system. Renormalization again reduces the differences between exact and correlated Coulomb integrals. The spikes associated with integrals that differ the most involve molecular orbitals in the virtual space and as will be shown in the next section have negligible effect on the excitation spectrum.

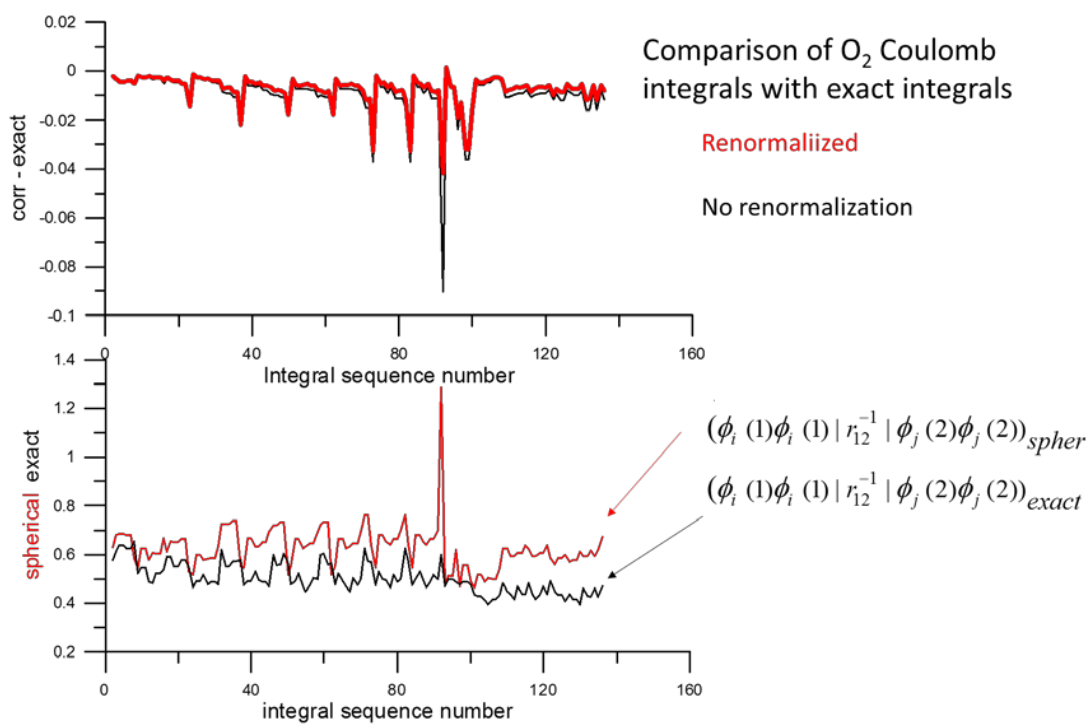

**Fig. 1. Comparison of integrals for $O_2$.** The upper graph shows the difference between the correlated and the exact value of the Coulomb integral over molecular orbitals, $(\varphi_i(1)\varphi_i(1)|r_{12}^{-1}|\varphi_j(2)\varphi_j(2))$ $i \leq j$, for each integral, 1-136, with and without renormalization In the lower graph, the value of the Coulomb integral and the value calculated using the spherical component approximation are compared.

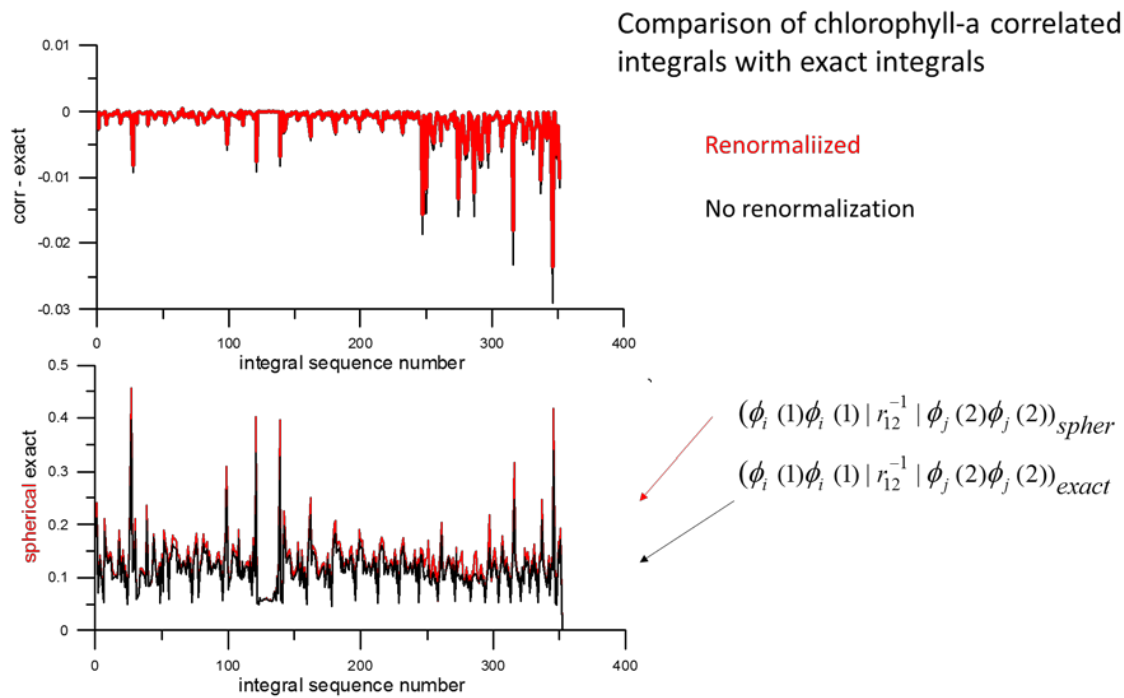

**Fig. 2. Comparison of integrals for chlorophyll-a.** The upper graph shows the difference between the correlated and the exact value of the Coulomb integral over molecular orbitals, $(\varphi_i(1)\varphi_i(1)|r_{12}^{-1}|\varphi_j(2)\varphi_j(2))$ $i \leq j$, for selected integrals, see text, with and without renormalization  In the lower graph, the value of the Coulomb integral and the value calculated using the spherical component approximation are compared. The agreement is better in this delocalized system than in $O_2$.

**Excited states of ethylene and chlorophyll-a**

     We consider in this section the application of the density component polarization method to excited electronic states of ethylene and chlorophyll-a. Both molecules have singlet π→π* excited states. In ethylene, there are geometry changes in the excited state and for the vertical or Franck Condon transition there are uncertainties about the spatial diffuseness of the excited singlet state. For chlorophyll-a, there are vibrational effects on the excitation spectrum and for the lowest excited singlet state, it is difficult to obtain an excitation energy as low as observed experimentally by theoretical methods that treat the ground and excited states at comparable accuracy. There is an extensive literature on both molecules. The present work is limited in scope and restricted to consideration of the Franck-Condon excitation and the question of whether the correlation method that was found promising for the diatomic molecule test cases is useful in describing these excited states. Both problems are treated by basis sets of comparable quality to those employed for the diatomic molecule studies and in the case of ethylene include additional diffuse p-type basis functions. The basis sets do not contain d-type atomic orbitals

that would normally be needed to correlate ionic distributions p(1)p(2) on the same nucleus. The chlorophyll molecule contains 50 Mg, C, N or O nuclei and the s,p basis totals 532 functions. While it is not difficult to add ~250 additional d-type basis functions to single-determinant SCF treatment, it is challenging to include these additional functions in a corresponding CI calculation.

In Table 5, energies for the ground and excited states of ethylene are reported. As noted earlier, there are no parameters in the method and the $E_\delta$ formula is the same as in the atomic and molecular calculations discussed earlier. Molecular orbitals are obtained from a SCF calculation on the triplet excited state (no $E_\delta$ contribution). The resulting molecular orbitals are used in a variational CI calculation containing ~30,000 determinants none of which involve d-orbital contributions.[8] The same molecular orbitals are used to investigate the effect of including the $E_\delta$ correlation contribution. A proper treatment of spin is necessary to distinguish between the triplet and singlet excited states and this is automatically correct in multi-determinant CI calculations. However, when the additional $E_\delta$ contribution is included, we want to maintain degeneracy of the $m_s = -1, 0, +1$ states for S=1 and to describe correctly the S=0, $m_s$=0 state. We accomplish this by adding $E_\delta$ (which is negative) to the Coulomb interaction (aa||bb) between singly occupied orbitals a and b and adding $E_\delta$ to the corresponding exchange interaction (ab||ab). The net effect is no $E_\delta$ correlation contribution to electrons in molecular orbitals with the same spin, but a correlation lowering of the Coulomb repulsion between electrons in orbitals with opposite spin.

**Table 5. Excited States of Ethylene**

Energies for SCF and CI calculations and for calculations that include $E_\delta$.

|  |  | 1-det[a] | CI | excitation (eV) |
|---|---|---|---|---|
| SCF, CI (no $E_\delta$) | gnd | -78.0075 | -78.2494 |  |
|  | $^3(\pi \to \pi^*)$ | -77.8922 | -78.0877 | 4.40 |
|  | $^1(\pi \to \pi^*)$ | -77.6452 | -77.9564 | 7.97 |
| $E_\delta$ included (no renorm) | gnd | -78.1913 | -78.4337 |  |
|  | $^3(\pi \to \pi^*)$ | -78.0698 | -78.2683 | 4.50 |
|  | $^1(\pi \to \pi^*)$ | -77.8418 | -78.1468 | 7.81 |
| $E_\delta$ included (renorm) | gnd | -78.1717 | -78.4160 |  |
|  | $^3(\pi \to \pi^*)$ | -78.0514 | -78.2510 | 4.49 |
|  | $^1(\pi \to \pi^*)$ | -77.8201 | -78.1270 | 7.86 |
| Expt | $^1(\pi \to \pi^*)$ |  |  | 7.65 |

[a] Total energies are in hartrees, 1 a.u. = 27.21 eV

The results in Table 5, show considerable improvement in the singlet π→π* excitation energy on adding $E_\delta$ while the triplet π→π* excitation energy which is nearly correct in the CI calculation without $E_\delta$ is less affected.

In Table 6, calculated energies for the ground and excited states of chlorophyll-a are reported. Spin states are treated the same as in ethylene. In chlorophyll, there are two single π→π* excitations that mix in each of the low-lying states. Configuration interaction is needed to account for this mixing, but, as seen in the Table, the excitation energies to the lower two singlet states are not in good agreement with experiment, lying 0.3 - 0.4eV too high.. The triplet-state excitation is not resolved experimentally. The table also shows that the two methods of including $E_\delta$ are identical that the excitation energies are not improved over the CI calculation without $E_\delta$. The table also shows that the simplest method in which only the few determinants needed to define the states are correlated by $E_\delta$ and the CI lowering is taken from the "exact" CI calculation is slightly better than the treatments in which $E_\delta$ is included in all determinants of the CI expansion. The change in excitation energy on adding $E_\delta$ is much less than found for the spatially more compact ethylene molecule. This result might be anticipated from the plots in Figure 2 that show smaller $E_\delta$ contributions and minor effects on renormalization in the delocalized molecular orbitals of chlorophyll. For chlorophyll, there are only 88 active molecular orbitals (42 electrons) and the CI calculation is unlikely to have reached its limit in accuracy.

**Table 6. Excited States of Chlorophyll-a**

Energies are given for SCF and CI calculations and for calculations that include $E_\delta$.
The first column of calculated excitation energies uses $E_\delta$ in all determinants while in the second
column only the leading determinants that define the state are correlated by $E_\delta$ and the CI lowering
is taken from the CI calculation with no $E_\delta$.

|  |  | 1-det[a] | CI | excitation energy (eV) | |
|---|---|---|---|---|---|
| SCF, CI (no $E_\delta$) | gnd | -2369.6360 | -2369.8716 | | |
|  | $^3(\pi\to\pi^*)$ | -2369.5673 | -2369.8092 | 1.70 | |
|  | $^1(\pi\to\pi^*)$ | -2369.5080 | -2369.7922 | 2.16 | |
|  | $^1(\pi\to\pi^*)$ | -2369.4755 | -2369.7729 | 2.68 | |
| $E_\delta$ included (no renorm) | gnd | -2369.8627 | -2370.0972 | | |
|  | $^3(\pi\to\pi^*)$ | -2369.7941 | -2370.0354 | 1.68 | 1.67 |
|  | $^1(\pi\to\pi^*)$ | -2369.7370 | -2370.0183 | 2.15 | 2.09 |
|  | $^1(\pi\to\pi^*)$ | -2369.7016 | -2369.9982 | 2.69 | 2.69 |
| $E_\delta$ included (renorm) | gnd | -2369.8461 | -2370.0807 | | |
|  | $^3(\pi\to\pi^*)$ | -2369.7776 | -2370.0190 | 1.68 | 1.67 |
|  | $^1(\pi\to\pi^*)$ | -2369.7204 | -2370.0019 | 2.14 | 2.10 |
|  | $^1(\pi\to\pi^*)$ | -2369.6852 | -2369.9819 | 2.69 | 2.68 |
| Expt | $^1(\pi\to\pi^*)$ | | | | 1.85 - 1.91[b] |
|  | $^1(\pi\to\pi^*)$ | | | | 2.14 - 2.23 |

[a] Total energies are in hartrees, 1 a.u. = 27.21 eV
[b] Ref. 16.

## Conclusion

1. An ansatz is proposed for introducing correlation effects that occur in electronic structure calculations by allowing basis function components of the interacting densities to polarize dynamically, thereby reducing the magnitude of the interaction. The modified Coulomb interactions are used in single-determinant or configuration interaction calculations. Exchange and other integrals over molecular orbitals are not modified.

2. The method partly accounts for dynamical correlation effects without explicitly introducing higher spherical harmonic functions into the molecular orbital basis. Molecular orbital densities are decomposed into a distribution of spherical components that conserve the charge and each of the interacting components is considered as a two-electron wavefunction acted on by an average field Hamiltonian plus $r_{12}^{-1}$. A method of avoiding redundancy and factors such as renormalization that improve the accuracy of the method are discussed.

3. Applications to atoms, negative ions and molecules representing different types of bonding and spin states show remarkably high accuracy and consistency.

4. Further studies are necessary to address higher order interference effects, self-consistent blocking and to go beyond the spherical component approximation.

## Acknowledgments

The author would like to acknowledge helpful discussions with Professor Mike Whangbo and his encouragement of this work. Support during the early stages of this work by the U.S. Department of Energy is gratefully acknowledged.